\documentclass{article}
\usepackage{ijcai22}

\usepackage{times}
\usepackage{soul}
\usepackage{url}
\usepackage[hidelinks]{hyperref}
\usepackage[utf8]{inputenc}
\usepackage[small]{caption}
\usepackage{graphicx}
\usepackage{amsmath}
\usepackage{amsthm}
\usepackage{booktabs}
\usepackage{algorithm}
\usepackage{algorithmic}
\usepackage{multirow}
\usepackage{color}
\usepackage{subfigure}
\usepackage[version=3]{mhchem}
\usepackage{stfloats}

\makeatletter
\newcommand{\printfnsymbol}[1]{%
  \textsuperscript{\@fnsymbol{#1}}%
}
\makeatother

\title{S2SNet: A Pretrained Neural Network for Superconductivity Discovery}

\author{
Ke Liu$^{1}$\footnotemark[1]
\and
Kaifan Yang$^{1}$\footnotemark[1]\and
Jiahong Zhang$^{2}$\And
Renjun Xu$^1$\footnotemark[2]
\affiliations
$^1$Zhejiang University\\
$^2$Yunnan University
\emails
\{rux, lk2017\}@zju.edu.cn
}

\begin{document}

\maketitle
\renewcommand{\thefootnote}{\fnsymbol{footnote}} 
\footnotetext[1]{These authors contributed equally to this work.} 
\footnotetext[2]{Corresponding author.} 

\begin{abstract}
Superconductivity allows electrical current to flow without any energy loss, and thus making solids superconducting is a grand goal of physics, material science, and electrical engineering. More than 16 Nobel Laureates have been awarded for their contribution in superconductivity research. Superconductors are valuable for sustainable development goals (SDGs), such as climate change mitigation, affordable and clean energy, industry, innovation and infrastructure, and so on. However, a unified physics theory explaining all superconductivity mechanism is still unknown. It is believed that superconductivity is microscopically due to not only molecular compositions but also the geometric crystal structure.
Hence a new dataset, S2S, containing both crystal structures and superconducting critical temperature, is built upon SuperCon and Material Project. Based on this new dataset, we propose a novel model, S2SNet, which utilizes the attention mechanism for superconductivity prediction. 
To overcome the shortage of data, S2SNet is pre-trained on the whole Material Project dataset with Masked-Language Modeling (MLM).
S2SNet makes a new state-of-the-art, with out-of-sample accuracy of 92\% and Area Under Curve (AUC) of 0.92.
To the best of our knowledge, S2SNet is the first work to predict superconductivity with only information of crystal structures. This work is beneficial to superconductivity discovery and further SDGs. Code and datasets are available in \href{https://github.com/zjuKeLiu/S2SNet}{https://github.com/zjuKeLiu/S2SNet} 
\end{abstract}

\section{Introduction}
Superconductors are crucial for sustainable everyday applications and attractive for the United Nations’ sustainable development goals (SDGs) \cite{miryala2021expert}. Superconductors have zero resistance, which allows perfect current transmission without losses. With superconducting cables, a continuous supply of power can be obtained for a timely availability of irrigation water in remote areas, which is a key to sustainability in performing agriculture. Then hunger can be effectively alleviated since it is partly caused by the lack of sustainability in performing agriculture. Affordable and clean energy can also be obtained because DC energy from places where power is plentiful, like solar plants in Africa (365 sunny days) can be delivered worldwide without losses. Zero resistance means ``no heat-generation and no energy-loss”, which is the key contribution in the climate action. For good health and well-being, superconducting magnets, which can produce magnetic field of high-quality, high-intensity, are well used as high-field magnets in magnetic resonance imaging, nuclear magnetic resonance, water purification, magnetic drug delivery, and so on.

Superconductivity has been a hot topic along many fields for its great significance to human beings and SDGs. However, the theory of superconductivity is still a puzzle till now. Benefiting from massive publicly available datasets in the area of material science, many machine learning methods have been applied to predicting material properties, generating structures, and so on \cite{schmidt2019recent}. It also sheds a light to modeling superconductors with machine learning methods. Much of the recent interest in superconductivity modelling has focused on critical temperature regression using measured properties of the material \cite{stanev2018machine,hamidieh2018data}, which is computationally expensive since feature engineering is required.

In AI for physical science, a lot of efforts have been made to model molecules with Graph Neural Networks (GNN) \cite{gnn1,gnn2} since the molecules are graphs naturally. However, quite few methods have been proposed for crystal modelling \cite{chen2019graph,park2020developing,xie2018crystal}.
Due to the different annotations among different datasets, the current supervised learning models can only be trained on small datasets, while other unlabeled data are not fully utilized.

Recently, attention mechanism is widely used in Natural Language Processing (NLP) \cite{attention_nlp} and Computer Vision (CV) \cite{dosovitskiy2020image} to model the interaction of contexts. The pre-training task, Masked-Language Modeling (MLM) task, in both NLP and CV, has shown great potential to get model better and more robust with astounding experimental results \cite{nlp_bert,attention_nlp}. Besides, with pre-training tasks, the unlabeled data can be utilized properly.

According to BCS theory, superconductivity can be microscopically explained as the interaction between electrons and lattice vibrations (phonons) \cite{bardeen1957microscopic}. Therefore, we propose a machine learning model for the way from \textbf{S}tructure \textbf{to} \textbf{S}uperconductivity (\textbf{S2SNet}), in which only crystal structures are required, to model the interaction in order to predict potential superconducting. Attention mechanism is applied to crystal structures. S2SNet is pre-trained on a large amount of unlabeled data to learn the interaction between different atoms in lattice.

Through S2SNet, we try to model the potential interaction between lattice dynamics and cooper pairs to predict the potential superconductor properties of material. It is experimentally verified that S2SNet outperforms other existing models. To the best of our knowledge, S2SNet is the first work showing that it is possible to predict the superconductivity of materials via their crystal structures. 

The main contributions of our work can be summarized as follows:
\begin{itemize}
	\item We propose a novel neural network to discover potential superconductivity of materials given only information of crystal structures  for the first time.
	
	\item A new dataset S2S containing both crystal structure and superconductivity critical temperature is built for the first time. It provides a benchmark for future machine learning based superconductor discovery.
	
	\item To provide insight and guide the superconductivity research for physicists, we present a list of potential superconductors for future experimental validation.
	
\end{itemize}

\section{Related Work}
\paragraph{Superconductivity Prediction.} Previous machine learning methods focus on predicting the superconductivity through the measured properties of materials, while these properties are obtained through complex and expensive feature engineering. The averaged valence-electron numbers, orbital radii differences, and metallic electronegativity differences were three critical features used in \cite{villars1988quantum}. Materials Agnostic Platform for Informatics and Exploration (Magpie) and AFLOW were used for feature engineering in \cite{stanev2018machine,hamidieh2018data}.

\paragraph{Crystal Modeling.} Quite few methods are proposed for crystal modeling. By connecting the atoms within a distance threshold, crystals were modeled as a graph with atoms as nodes in CGCNN  \cite{park2020developing,xie2018crystal}. External environmental conditions are taken into consideration by taking it as an additional nodes and connecting it with all the other nodes in MEGNet \cite{chen2019graph}.

\paragraph{Unsupervised Learning.} Various approaches have been proposed to make full use of the unlabeled data, most of which pre-train a model with the unlabeled dataset and finetune it on specific labeled dataset \cite{han2021pre}. Contrastive learning is commonly used for pre-training in CV \cite{le2020contrastive}. As pre-training tasks in NLP, diverse techniques such as mask language model (MLM), replaced token detection (RTD), and so on are often employed \cite{vaswani2017attention}. Following the pre-training model Bert, Zhang et. al. proposed ``MG-Bert" for molecule modelling with unsupervised learning \cite{zhang2021mg}.
\begin{figure}[tbp]
	\includegraphics[width=\columnwidth]{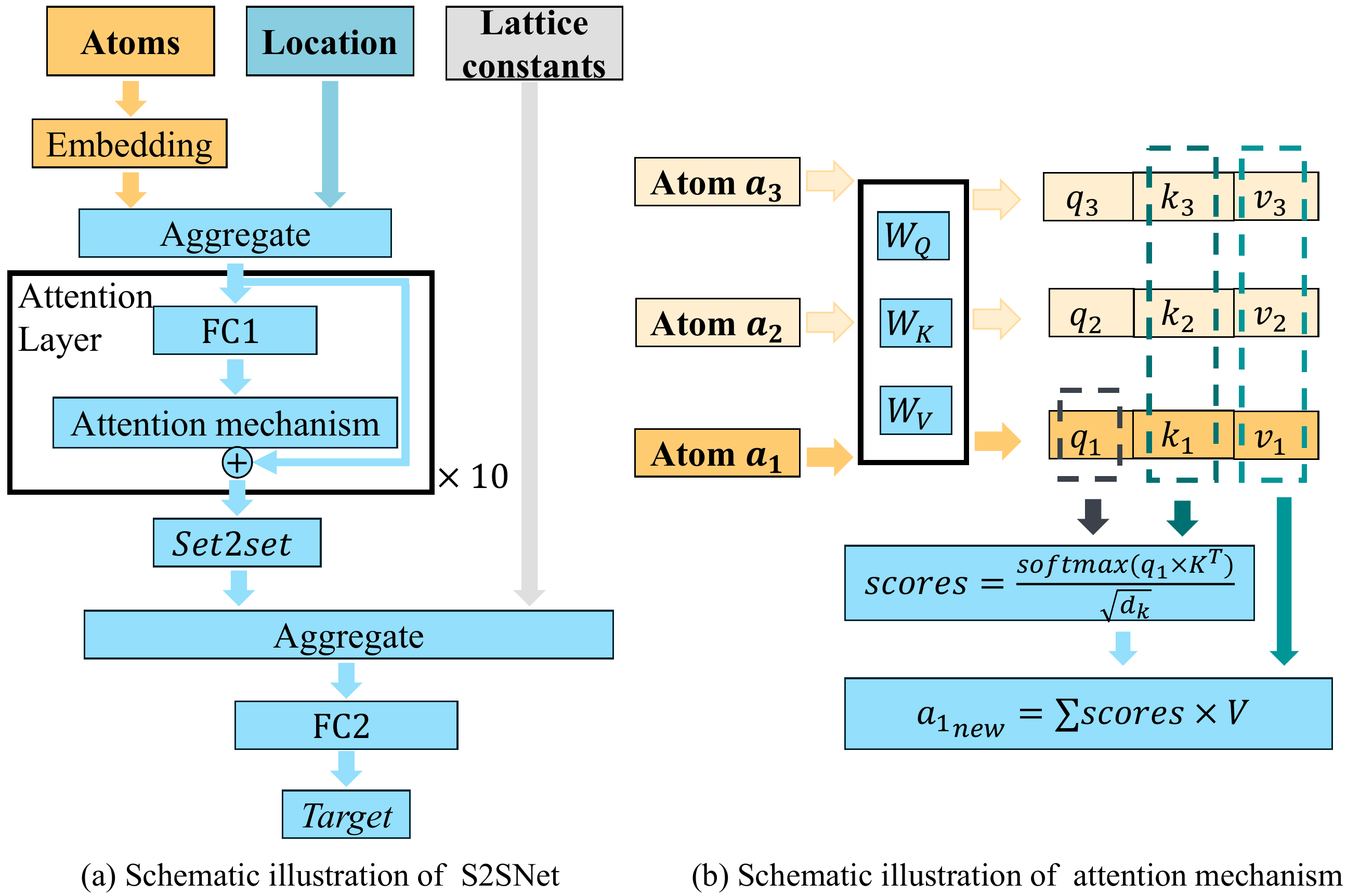}
	\caption{Illustration of S2SNet and attention mechanism. (a) Overall illustration of S2SNet. The Attention Module are performed for 10 times. $FC1$ and $FC2$ are two fully connected hidden layers. $FC2$ outputs the target $\hat{y}_t$. (b) Illustration of the Attention mechanism. The computation of representation of Atom $a_1$ is presented. Best view in color.
	}\label{fig:model}
\end{figure}

\section{Methodology}
Attention mechanism is proposed in the filed of NLP and widely used in CV recently \cite{vaswani2017attention,dosovitskiy2020image}. It is applied to the crystals in this study.
\subsection{S2SNet}
Fig.~\ref{fig:model} shows the overall illustration of S2SNet and the attention mechanism. 
In this paper, it is defined that a crystal $c\in C$ consists of atoms $A=\{a_1,a_2,\cdots,a_n\}$, the location of atoms $P=\{p_1,p_2,\cdots,p_n \}$, and the lattice parameters $U=\{u_1,u_2,u_3\}$, i.e. $c = \{A, P, U\}$. $n$, $a$, $p$ and $U$ denote the number of atoms in a unit cell, the atomic number, the actual 3-D location of an atom in the crystal, and the physical dimension of a unit cell respectively.
The overall workflow of S2SNet can be described as followings:
\begin{enumerate}
	\item Atoms $A$ are embedded into 512-dim vectors through the Embedding layer. Expanded distance with Gaussian basis centered at 512 points linearly placed between 0 and 5.12 \r{A} is applied to the location $P$ of atoms in the crystal \cite{chen2019graph}. Specifically, the Euclidean distances $P'$ between atoms and the center of crystal i.e. the mean coordinates of all atoms, are calculated. Then Gaussian basis are performed as Eq.~\ref{equ:distance}, where $P'_0$ are 512 distances from 0 to 5.12 \r{A} and $\sigma=2$ \cite{chen2019graph}.
	\begin{equation}\label{equ:distance}
	\begin{aligned}
	A' &= embedding(A), \\
	\hat{P} &= e^{-\frac{(P'-P'_0)^2}{\sigma^2}} 
	\end{aligned}
	\end{equation}
	
	\item The embedding of atoms $A'$ are atom-wise aggregated with the distance $\hat{P}$ as Eq.~\ref{equ:agg}, where $W_\alpha$ is a matrix to get the position $\hat{P}$ into the same dimension as $A'$ and model the interaction between different positions.
	\begin{equation}\label{equ:agg}
	A'' = A' + W_\alpha \cdot \hat{P}
	\end{equation}

	\item Attention mechanism is performed on each atom respectively with its neighborhood atoms. Thus the embedding of atom contains all the information of itself and neighbors. $neighbor(a'')$ is defined as the atoms within the Euclidean distance, $\rho$, from $a''$. In this paper, $\rho$ equals to 2 \r{A} according to \cite{chen2019graph}.
	\begin{equation}\label{equ:attention}
	a_i''' = Attention(a_i'', neighbor(a_i''))
	\end{equation}

	\item To aggregate the embedding of all atoms in crystal, the $Set2Set$ \cite{2015arXiv151106391V} is performed on the set $A'''$.
	\begin{equation}\label{equ:set2set}
	b = Set2Set(A''')
	\end{equation}

	\item Lattice parameters are aggregated to get the final representation of crystal as Eq.~\ref{equ:agg2}, where $W_\theta$ is a matrix to get the lattice parameters $U$ into the same dimension as $b$.
	\begin{equation}\label{equ:agg2}
	b' =  b + W_\theta \cdot U
	\end{equation}

	\item Through the $FC2$ layer, target can be obtained and loss function is binary cross entropy as Eq.~\ref{equ:binary}, where $y_t$ is the ground truth of whether it is a superconductor and $\hat{y}_t$ is the prediction of S2SNet.
	\begin{equation}\label{equ:binary}
	\mathcal{L}_{finetune} = -\frac{1}{2} [ y_t \log \hat{y}_{t} + (1- y_t) \log (1-\hat{y}_{t})]
	\end{equation}
	
\end{enumerate}

\begin{figure}[tbp]
	\centering
	\includegraphics[width=1\columnwidth]{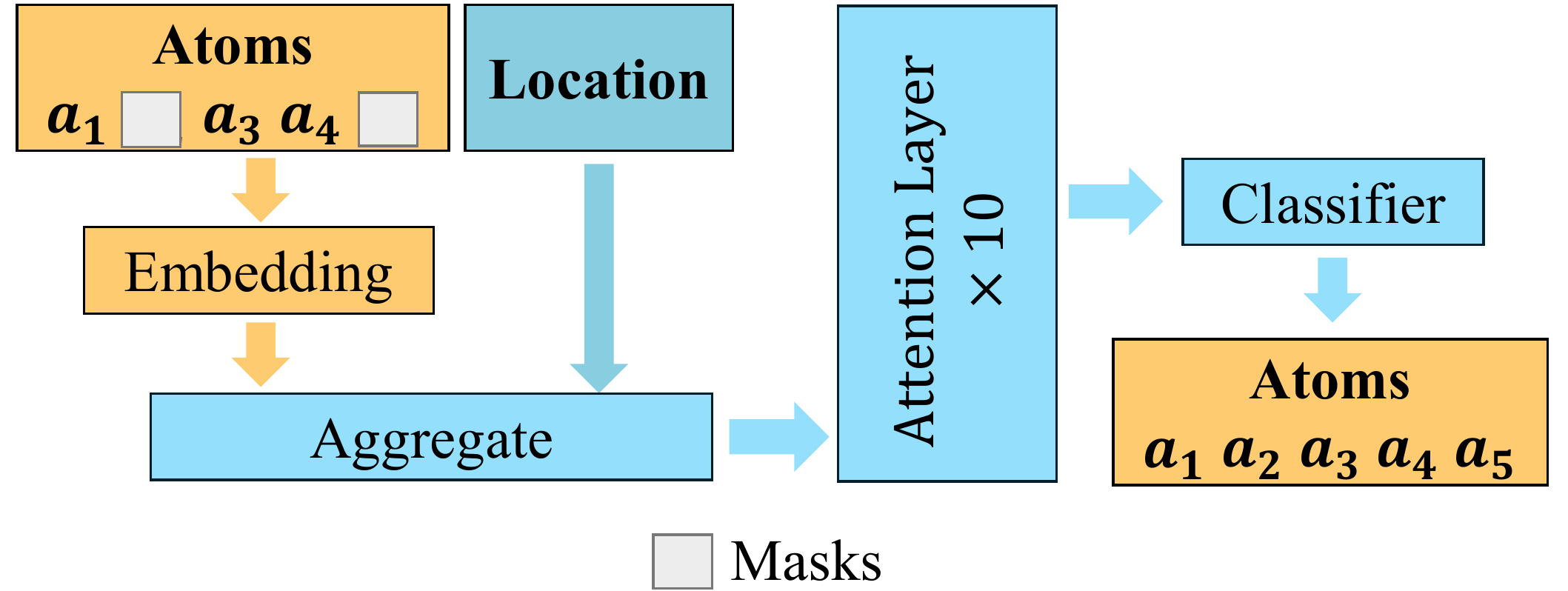}
	\caption{Illustration of MLM pre-training
		task. Classifier contains a layer with 95 neurons and a $softmax$ layer and outputs $N$ vectors of 95-dimension $\hat{y}$.}\label{fig:pre-train}
\end{figure}

In Fig.~\ref{fig:model}(b), an example of the attention mechanism is illustrated. Suppose that the neighbors of atom $a_1$ are $a_2$ and $a_3$. To get the representation of atom $a_1$, $a_1$ and its neighbors, $a_2$ and $a_3$, are firstly multiplied with $W_Q$, $W_K$ and $W_V$ respectively. Three different groups of $\{q, k, v \}$ are obtained. $q_1$ from $a_1$ is multiplied with $k_1, k_2, k_3$ from all the atoms to get the $score$ as Eq.~\ref{equ:score}, where $d_k$ is the dimension of $k$. Finally, the representation of atom $a_1$ can be obtained as the weighted sum of $V$ from all the atoms as Eq.~\ref{equ:final}, where $N$ is the number of $a_1$'s neighbors plus one. Representations of other atoms $a_i \in A$ can also be obtained like this.

\begin{equation}\label{equ:score}
scores = \frac{softmax(q_1 \cdot K^T)}{\sqrt{d_k}}
\end{equation}

\begin{equation}\label{equ:final}
a_1''' = \sum_{i=1}^N scores_i \cdot v_i
\end{equation}

\subsection{MLM Pre-training Task}
MLM pre-training task is a self-supervised learning method to model the interaction between atoms via classification as Fig.~\ref{fig:pre-train}. In this task, 15\% (at least 2) of the atoms in a crystal are masked, and the model are trained to predict the class of masked atoms. Therefore no extra labels except the structure of the crystal are required. Since there are 95 elements in S2S dataset, this pre-training is a 95-classification task. The objective of this pre-training can be defined as Eq.~\ref{equ:loss1}, where $N$ and $95$ denote the number of masked atoms and the number of element classes respectively. $y_{i,j}$ is the ground truth and $\hat{y}_{i,j}$ is the prediction of S2SNet for ``atom $i$ is the element $j$", 0 for False and 1 for True.

\begin{equation}\label{equ:loss1}
\mathcal{L}_{pre-train} = - \frac{1}{N}\sum_{i=1}^{N} \sum_{j=1}^{95}y_{i,j} \log \hat{y}_{i,j}
\end{equation}

After convergence of training, the \textit{Embedding}, \textit{Aggregate}, and \textit{Attention} layers are cloned. Finally, S2SNet is then fine-tuned on S2S dataset with the pre-trained layers frozen.

\section{Experiments}

To verify S2SNet, a dataset of superconductors is constructed, on which extensive experiments are conducted. Compared with other models, the experiments demonstrate that S2SNet effectively captures the crystal structural information for superconductivity prediction.

\begin{table}[btp]
\centering 
 \begin{tabular}{cccc}
     \hline
      Dataset & $\left| U\right|$  &  $Max\left(\left| \boldsymbol{A}\right|\right)$ &  $\left| V\right|$ \\
       \hline
       S2S & 1,685 & 248 & 95\\ 
       pre-train  & 69,239 & 444 & 95\\
       \hline
   \end{tabular}
\caption{Statistics of the datasets. $\left| V\right|$ indicates the number of classes of atoms. $\left| U\right|$ indicates the number of materials} 
\label{dataset_info}
\end{table}

\subsection{Setup}
\paragraph{S2S Dataset.} We collect a total of 69,239 crystal structures from Material Project database \cite{mp}, each of which contains lattice structure, formation energy, and band gap of crystals. 33,810 superconductors and their critical temperatures are obtained from SuperCon\footnote{https://supercon.nims.go.jp/en/}, which stores the critical temperature of substances. In addition, this paper also obtains structural information of part of materials from databases such as OQMD \cite{oqmd}, and ICSD \cite{icsd}.
The information in the databases above are used to construct S2S dataset with 1,685 crystals, which includes 421 currently known crystal superconductors and other 1,264 materials without known superconductivity sampled from Material Project database. Each entry includes the name of the substance, its crystal structure(atoms, location of atoms and lattice parameters), and whether it is a superconductor. This statistics of S2S dataset and pre-training dataset are shown in Table~\ref{dataset_info}.

\paragraph{Compared Approaches.} We mainly compare our proposal with several superconductivity prediction methods, \textbf{MEGNet} \cite{chen2019graph}, \textbf{Statistic} \cite{stanev2018machine}, and \textbf{CGCNN} \cite{xie2018crystal}. MEGNet uses graph networks to model molecules and crystals universally for property prediction. Statistic achieves a pleasant result on predicting superconductor which calculates 145 properties of materials with MAGPIE \cite{baldow2017magpie} and put these features into a random forest for training. CGCNN is specifically designed for accurate and interpretable prediction of crystal material properties where convolutional neural networks are performed on crystal graph.

\paragraph{Evaluation Metrics.} We mainly employ Accuracy (\textbf{ACC}) and Area Under Curve (\textbf{AUC}) for superconductor prediction. In addition, we also use visualization and interpretability analysis to verify our S2SNet.

\subsection{Experimental Results and Discussion}

\begin{figure}[b!]
    \centering
	\includegraphics[width=1\columnwidth]{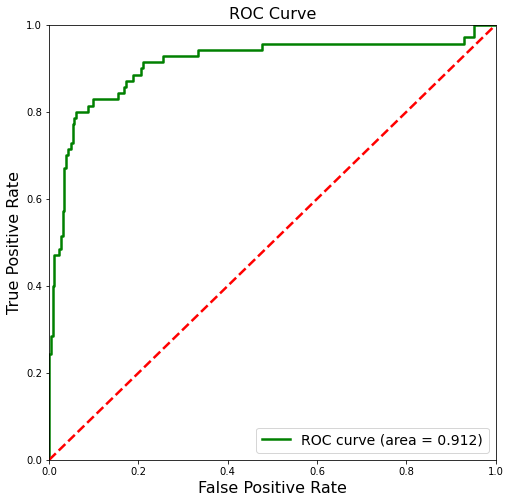}
	\caption{ROC curve for test data of 10 attention layer pre-trained S2SNet. The horizontal and vertical coordinates are FPR and TPR respectively.
	}\label{fig:roc}
\end{figure}

\begin{table}[t]
	\centering
	\begin{tabular}{c c c  c c}
		\toprule
		\multirow{2}{*}{Method} & \multicolumn{2}{c}{pre-train} & \multicolumn{2}{c}{without pre-train}\\
		~ & ACC(\%) & AUC(\%) & ACC(\%) & AUC(\%)\\
		\midrule
		MEGNet & 80.30 & 79.10 & 78.37 & 75.43 \\
		Statistic & -   &  -  &   83.33   &    78.52  \\
		CGCNN & 82.42 & 86.40 & 79.69 &  82.45 \\
		S2SNet(ours) &  \textbf{91.28}   &  \textbf{91.18}  &  \textbf{89.53}  &  \textbf{90.24}  \\
		\bottomrule
	\end{tabular}
	\caption{The experimental results of different models on the S2S dataset.}
	\label{tab:result1}
\end{table}

\paragraph{Comparison on S2S Dataset.}
All these models are trained and tested directly on S2S dataset. 
To bridge the shortage of data, MEGNet and CGCNN are pre-trained with the task of predicting formation energy following the paper \cite{chen2019graph} and S2SNet is pre-trained with the MLM task both on the Material Project database. 80\% of S2S dataset are randomly selected to train the models and the left 20\% are used to test the models. All the models are evaluated in the same way. Adam optimizer is used in S2SNet and the settings of other models follow their papers.

\begin{figure}[btp]
    \centering
	\includegraphics[width=\columnwidth]{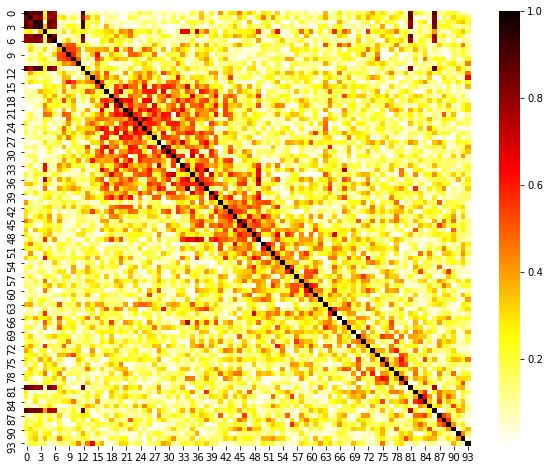}
	\caption{Pearson correlations between elemental embedding vectors. Elements are arranged in order of increasing Mendeleev number for easier visualization of trends. The number denotes the Mendeleev number.}\label{fig:pearson}
\end{figure}

The experimental results are shown in Table~\ref{tab:result1}. S2SNet consistently outperforms the other competitors both on accuracy (ACC) and Area Under Curve (AUC). For example, S2SNet gets about 6\% improvements on ACC than Statistic and about 7\% improvements on AUC than CGCNN without pre-training and almost 11\%/14\% improvements than MEGNet on ACC/AUC. With the pre-training task, all these three models, MEGNet, CGCNN and S2SNet are improved significantly. Since S2SNet is pre-trained on the unsupervised MLM task, it can be pre-trained on other much bigger dataset without any labels. However, MEGNet and CGCNN must be trained on the dataset with specific labels such as formation energy and so on. This also means that S2SNet can be further improved with larger dataset in the future work.

Fig.~\ref{fig:roc} shows the receiver operating characteristic (ROC) curve for the superconductor classifier. The overall test accuracy is 91.28\%, and the area under curve for the receiver operation conditions is 0.9118.

\begin{figure}[b!]
    \centering
	\includegraphics[width=1\columnwidth]{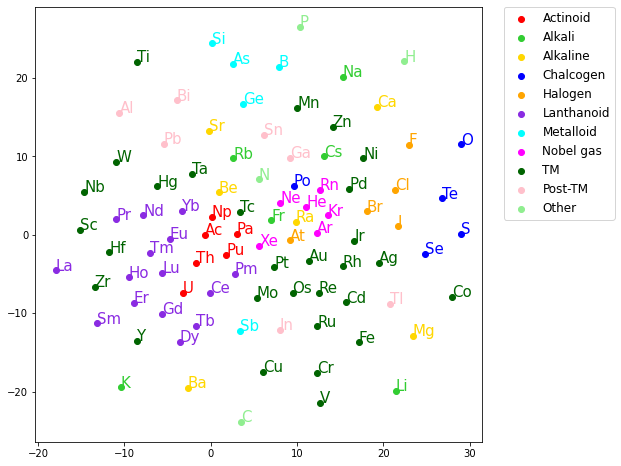}
	\caption{Elemental embeddings in 2-dimension with t-SNE.
	}\label{fig:element}
\end{figure}

\subsection{Ablation Study}

Ablation studies are also conducted to verify our S2SNet. Table~\ref{tab:result2} shows the experimental results of S2SNet with different number of attention layers on S2S dataset. Experiments demonstrate that more layers lead to better AUC. For both efficiency and accuracy, the number of attention layers in our S2SNet is 10.
\begin{table}[tbp]
	\centering
	\begin{tabular}{c  c  c  c}
		\toprule
		\# attention layers & 5 & 10 & 15 \\
		\midrule
		ACC(\%) & 89.83  & 91.28  & \textbf{92.31} \\
		AUC(\%) & 89.38  & 91.18  & \textbf{92.27} \\
		\bottomrule
	\end{tabular}
	\caption{The experimental results of models with different numbers of attention layers.}
	\label{tab:result2}
\end{table}

Further experiments are conducted to verify that the lattice parameters are helpful in S2SNet. Table~\ref{tab:result3} shows results of 10 attention layer S2SNet without the input of lattice parameters. With lattice parameters, models perform much better no matter pre-trained or not.

\begin{table}[tbp]
	\centering
	\begin{tabular}{c c c c c}
	    \toprule
	    \multirow{2}{*}{Method} & \multicolumn{2}{c}{pre-train} & \multicolumn{2}{c}{without pre-train}\\
	     & w & w/o & w & w/o \\
		\midrule
		ACC(\%) & \textbf{91.28} & 90.41 & \textbf{89.53} & 87.79 \\
		AUC(\%) & \textbf{91.18} & 89.22 & \textbf{90.24} & 86.20  \\
		\bottomrule
	\end{tabular}
	\caption{The experimental results of 10 attention layer models without the input of lattice parameters, where w and w/o denote the models with and without lattice parameters respectively.}
	\label{tab:result3}
\end{table}

\begin{figure}[b!]
    \centering
	\includegraphics[width=0.97\columnwidth]{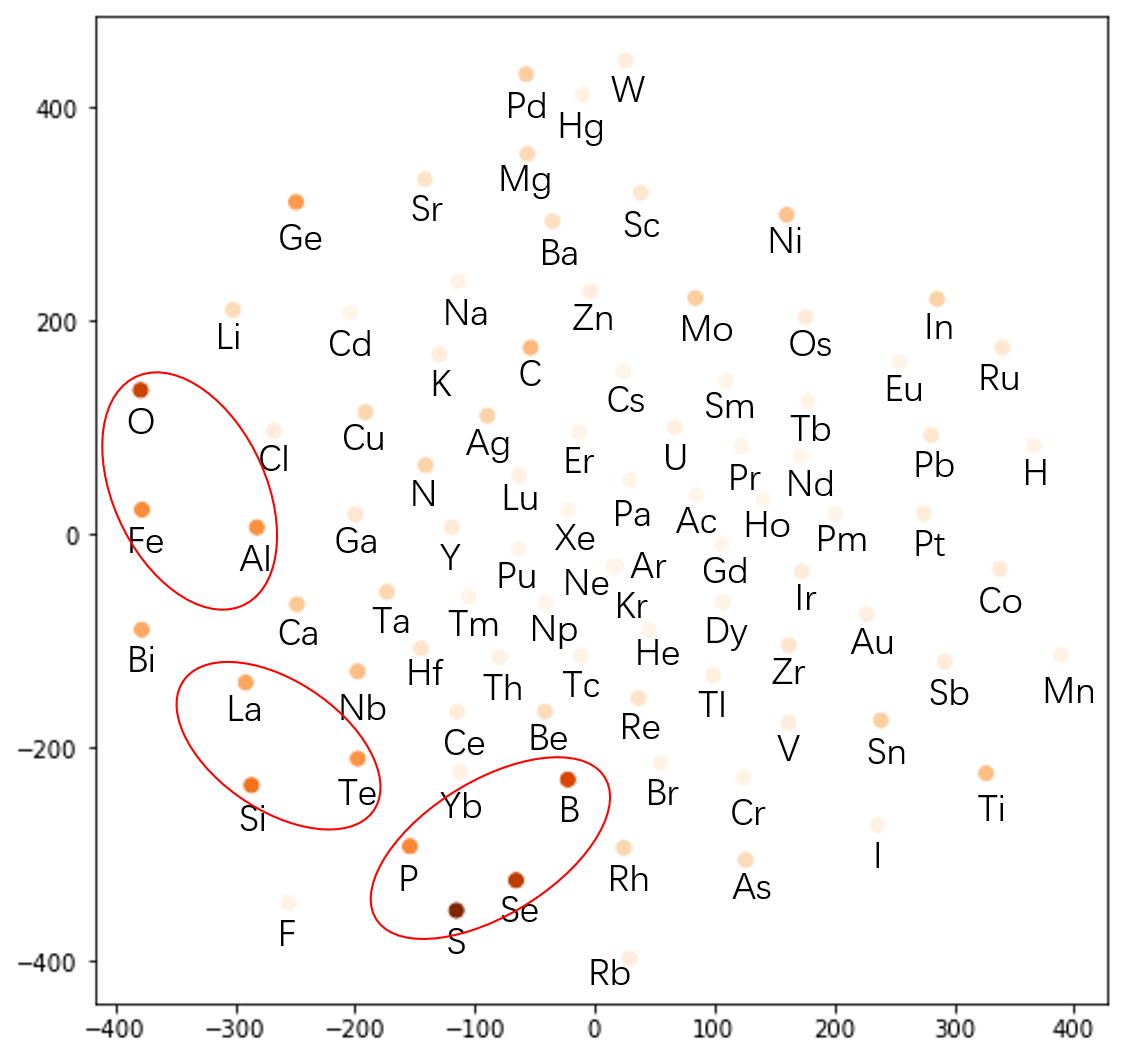}
	\caption{Elemental embeddings in 2-dimension with t-SNE. The darker the color, the more often the element appears in the materials that our model determines to be superconductors.
	}\label{fig:cluster}
\end{figure}

\subsection{Interpretability on S2SNet}
According to \cite{zhou2018learning}, the interpretability and reproducibility of known chemical common knowledge for machine learning models in material science is critical. To this end, the embeddings of all the element are extracted from S2SNet. The Pearson correlation coefficient between different elements are calculated as Eq.~\ref{equ:similarity}, where $m$ is the dimension of elemetal embedding vector. As is shown in Fig.~\ref{fig:pearson}, the Pearson correlation coefficient  between elemental embedding vectors follow the trend of the periodic table of elements. For example, the  chalcogen, halogen, lanthanoid, transition metals, metalloid, and actinoid show highest similarities within their groups. For better observation on the elemental embedding vectors, t-SNE\cite{hinton2002stochastic} is used to map the high-dimension embedding vectors into 2-dimension space in Fig.~\ref{fig:element}. Elements with similar properties are close to each other, such as the Actinoid, the Chalcogen, the Halogen and so on. All the visualizations show that S2SNet learns the common intuition in chemistry.

\begin{equation}\label{equ:similarity}
r_{ij} = \frac{\sum_{k=1}^m(a_{ik}-\overline{a}_i)(a_{jk}-\overline{a}_j)}{\sqrt{\sum_{k=1}^m(a_{ik}-\overline{a}_i)^2\sum_{k=1}^m(a_{jk}-\overline{a}_j)^2 }}
\end{equation}

\subsection{Potential Superconducting Materials}
In order to provide more insight to the field of superconducting materials research, all the materials we have are fed into S2SNet to predict whether they are superconductors or not. Frequency of elements in materials that our model determines to be superconductors is shown in Fig.~\ref{fig:static} in Appendix~\ref{appendix:ps}. Orange and blue ones denote the number of atoms and the number of materials that contain corresponding atom respectively. \ce{Si}, \ce{S}, \ce{B}, \ce{La}, \ce{Fe}, \ce{Se}, \ce{Ge}, \ce{P}, \ce{Al}, \ce{N}, \ce{O}, \ce{Bi}, \ce{C}, \ce{Ca}, \ce{Te}, \ce{Ti} and \ce{Ni} appear in more than 300 superconductors which, in our opinion, are supposed to be crucial elements for superconductor. And the known superconducting elements \ce{Nb}, \ce{Pb}, and so on are also identified as the crucial elements in S2SNet.
To verify that whether our model has learned these knowledge, we make the embedding layer learnable and train the overall model with the superconductivity prediction. Atomic embeddings are visualized by t-sne in Fig.~\ref{fig:cluster}. The atoms with high frequency are grouped together.

We get 5468 potential superconductors out of the 69239 materials, and the element combination of part of  representative potential superconductors are list in Table~\ref{tab:super} in Appendix~\ref{appendix:ps}. The existing Iron-based, Cuprate, Hydrogen-based and Metal superconductors are identified with high score by our model. Some potential superconductors with similar structures are also predicted. Besides, the other potential superconductors with extremely high score are also given for further experimental synthesis verification. The results are reasonable, since $Graphene$ and $SiC$, which are not in the training set, are predicted to be superconductors. The prediction accuracis for Iron-based, Cuprate, Hydrogen-based superconductors are 97.64\%, 92.00\% and 96.89\% respectively. It is worth mentioning that all the 21 Iron-based superconductors are correctly predicted.

\subsection{Why Are They Superconductors}

\begin{figure*}[t!]
  \centering 
  \subfigure[\ce{BaFe2S3}]{ 
    \label{fig:subfig:a} 
    \includegraphics[width=0.76\columnwidth]{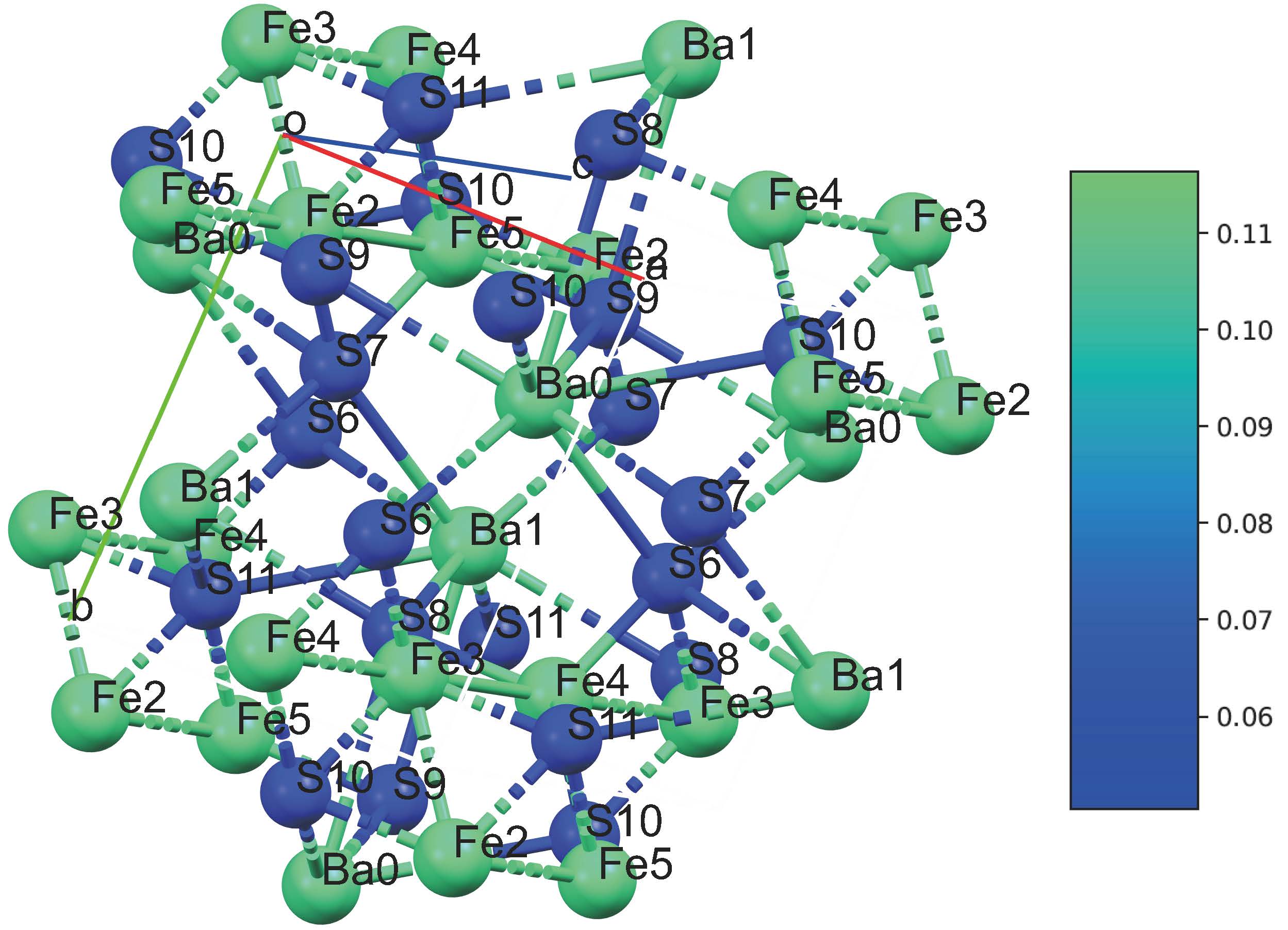}}
  \hspace{0.01\columnwidth}
  \subfigure[\ce{H3S}]{ 
    \label{fig:subfig:b} 
    \includegraphics[width=0.60\columnwidth]{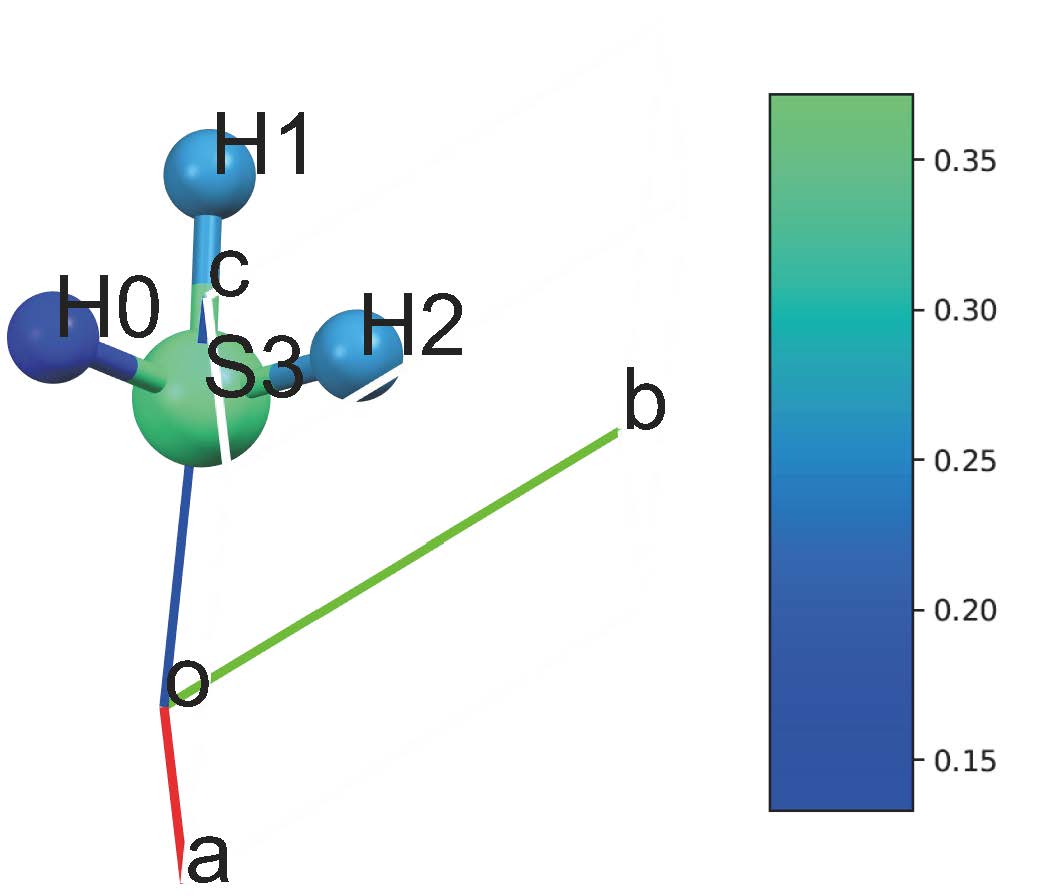}} 
  \hspace{0.01\columnwidth}
  \subfigure[\ce{Sr2CuClO2}]{ 
    \label{fig:subfig:c} 
    \includegraphics[width=0.62\columnwidth]{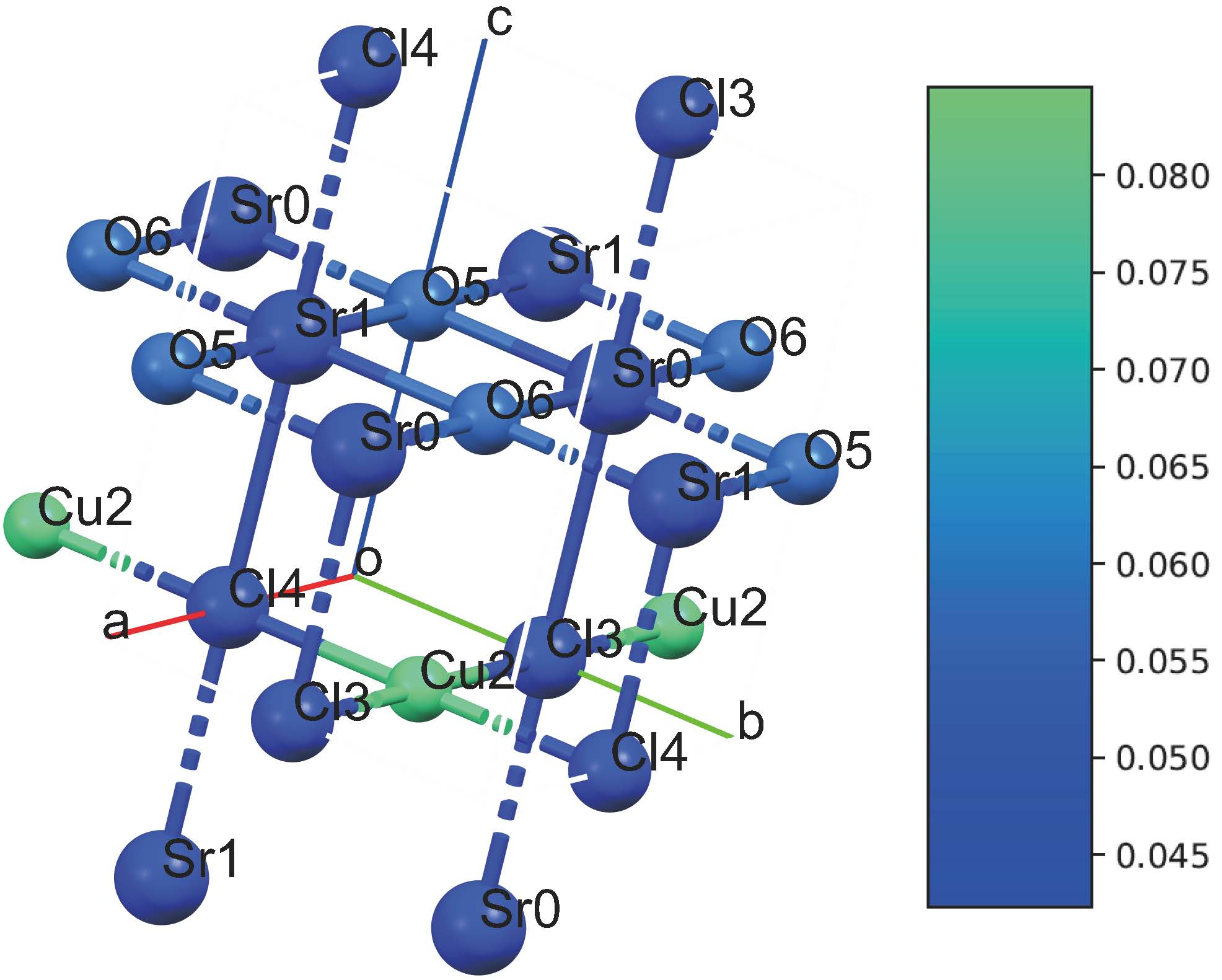}} 
  \caption{Contribution of each atom to superconductivity in \ce{BaFe2S3}, \ce{H3S}, and \ce{Sr2CuClO2}. Color denotes the weight. Best viewed in color}
  \label{fig:subfig} 
\end{figure*}

Class Activation Map (CAM) is used to explore how superconductivity comes \cite{kwasniewska2017deep}. CAM is a technique commonly used in CV for interpretability, i. e., highlighting the parts of a picture where the model identifies it as a specific class. Herein, we highlight the atoms in crystals that are critical for superconductivity with CAM. The experimental findings are consistent with the previously proposed theories. We take the \ce{BaFe2S3}, \ce{H3S}, and \ce{Sr2CuClO2} for example. The contribution of atoms in each crystal to superconductivity is shown in Fig.~\ref{fig:subfig} and the color denotes the contribution. \ce{BaFe2S3} in Fig.~\ref{fig:subfig:a} represents the Iron-based superconductors and the \ce{Fe} is more crucial for superconductivity. \ce{H3S} represents the Hydrogen-based superconductors. Fig.~\ref{fig:subfig:b} shows that both \ce{H} and \ce{S} are important for superconductivity and the three hydrogen atoms are of different importance. The hydrogen atoms that are closer to the \ce{S} is more crucial which is consistent to previous experimental results that with higher pressure, the atoms in a crystal are closer, which can cause the superconductivity. \ce{Sr2CuClO2} represents the superconductors based on Cuprate. Fig.~\ref{fig:subfig:c} demonstrate that \ce{Cu} and \ce{O} are crucial atoms.

\section{Conclusion}
A new dataset, S2S, is built as a benchmark for future machine learning based superconductivity discovery. Based on S2S, a novel pre-trained neural network, S2Snet, is built for  superconductivity prediction solely from their crystal structures. 
S2SNet leverages attention mechanism to model the interaction of atoms in the crystal cells. With the pre-training task, MLM, massive unlabeled data can be properly utilized for S2SNet to learn better representation of atoms under interactions. Indeed S2SNet outperforms other methods significantly no matter with or without pre-training. Furthermore, it is straightforward to extend S2SNet to predict other properties of crystals as well.
We expect that S2S dataset and S2SNet could inspire the machine learning community to make more impact on  AI for science and SDGs. Finally, the current dataset is still relatively small. Enriching S2S dataset will be our future work.

\section*{Appendix}
\appendix
\renewcommand\thefigure{\Alph{section}\arabic{figure}}
\renewcommand\thetable{\Alph{section}\arabic{table}}  
\renewcommand\theequation{A.\arabic{equation}}
\setcounter{equation}{0}
\setcounter{table}{0}
\setcounter{figure}{0}

\section{Potential Superconductors}
\label{appendix:ps}
\begin{figure}[t!]
    \centering
	\includegraphics[width=0.95\columnwidth]{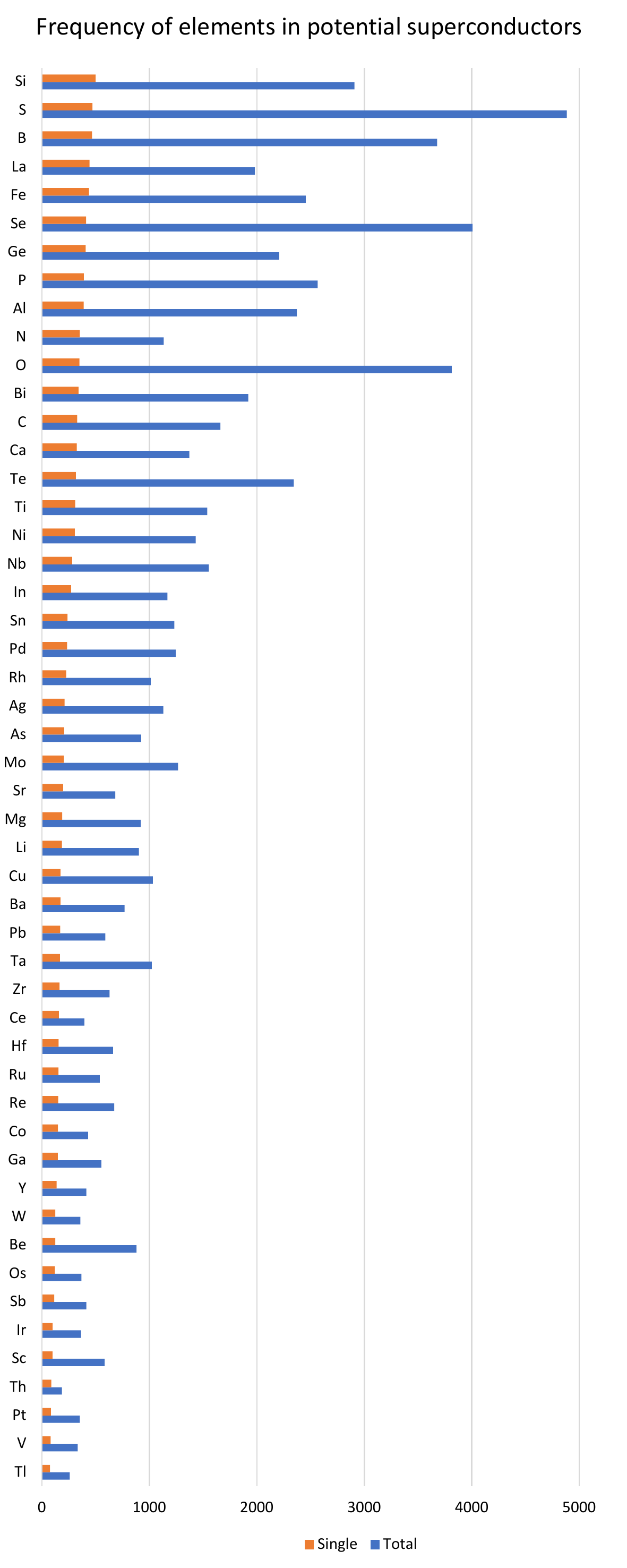}
	\caption{ Frequency of elements in materials that our model determines to be superconductors. Orange and blue ones denote the number of atoms and the number of materials that contain corresponding atom respectively.
	}\label{fig:static}
\end{figure}

Potential superconductors are shown in Table~\ref{tab:super}. Frequency of elements in materials that our model determines to be superconductors is shown in Fig.~\ref{fig:static}. Here, we predict whether a material has the potential to be superconducting, regardless of temperature and pressure.

\begin{table}[t!]
	\centering
	\begin{tabular}{c c c }
		\toprule
		~ & Element combination & Score\\
		\midrule
		\multirow{5}{*}{Iron-based} & LiFeAs* &	0.9890\\
		~ & InFeAs & 0.9939\\
		~ & TlFeSe2 & 0.9825\\
		~ & ZrFeSe & 0.9849\\
		~ & SmFeAsO	& 0.9890\\
		\midrule
		\multirow{5}{*}{Cuprate} & La2CuO4* & 0.9396 \\
		~ & LiCuO &	0.9738 \\
		~ & La7Ti27Al5Cu24O96 & 0.9999 \\
		~ & Ba9La9Cu9Ru9O46 & 0.9999 \\
		\midrule
		\multirow{5}{*}{Hydrogen-based}  & H2S* & 	0.6269 \\
		~ & Mo3H & 	0.9510 \\
		~ & Ti3SnH & 	0.9009\\
		~ & NbH2 & 	0.7034 \\
		~ & Ta2H & 	0.6826\\
		\midrule
		\multirow{2}{*}{Metal} & Hg* &	0.9953 \\
		~ & NbSn2* & 0.9976\\
		\midrule
		\multirow{27}{*}{Others} & Bi4Te3*	&	1	\\
		~&	C*	&	1	 \\
		~&	TaSe2	&	1 \\
		~&	Na3Sr14Nd3Ti20O60	&	1 \\
		~&	Ge4Te7As2	&	1\\
		~&	Ge1Bi2Te4	&	1 \\
		~&	Tl12In8Se20	&	1 \\
		~&	Ta4AgS8	&	1	\\
		~&	Ge3Bi2Te6	&	1 \\
		~&	Tl8In12Se20	&	1 \\
		~&	Ba5Bi14O21	&	1 \\
		~&	ErZr11Si11PO48	&	1 	\\
		~&	SiC	&	1 \\
		~&	TaS2	&	1	\\
		~&	PbI2	&	1 	\\
		~&	Bi8Te9	&	1 	\\
		~&	Sb16Te3	&	1 	\\
		~&	Al8C3N4	&	1 	\\
		~&	Sr8Ru7O24	&	1 	\\
		~&	Sb8Te3	&	1	\\
		~&	Sc2Te3	&	1	  \\
		~&	Zr3S4	&	1	 \\
		~&	La39Se56	&	1 	\\
		~&	Rb4Mo21Se24	&	1 	\\
		~&	Ti7S12	&	1 	\\
		\bottomrule
	\end{tabular}
	\caption{The element combination of potential superconductors. * denotes that the superconductor has already been discovered.}
	\label{tab:super}
\end{table}

\bibliographystyle{named}
\bibliography{ijcai22}

\end{document}